\documentclass[sn-mathphys,Numbered,iicol]{sn-jnl}

\usepackage[utf8]{inputenc}
\usepackage{graphicx}%
\usepackage{multirow}%
\usepackage{amsmath,amssymb,amsfonts}%
\usepackage{amsthm}%
\usepackage{mathrsfs}%
\usepackage[title]{appendix}%
\usepackage{xcolor}%
\usepackage{textcomp}%
\usepackage{manyfoot}%
\usepackage{braket}

\usepackage{geometry}
\newcommand{\nep}{\operatorname{e}}

\newcommand{\opcdag}[1]{{\hat{c}^{\dagger}}_{#1}}

\newcommand{\opc}[1]{{\hat{c}^{\phantom \dagger}}_{#1}}

\begin{document}

\title[The impact of different unravelings in a monitored system of free fermions]
      {The impact of different unravelings in a monitored system of free fermions}

\author[1]{\fnm{Giulia} \sur{Piccitto}}\email{giulia.piccitto@unict.it}

\author[2]{\fnm{Davide} \sur{Rossini}}\email{davide.rossini@unipi.it}

\author[3]{\fnm{Angelo} \sur{Russomanno}}\email{angelo.russomanno@unina.it}

\affil[1]{\orgdiv{Dipartimento di Matematica e Informatica}, \orgname{Universit\`a di Catania},
  \orgaddress{\street{Viale Andrea Doria 6}, \city{Catania}, \postcode{95125}, \country{Italy}}}

\affil[2]{\orgname{Dipartimento di Fisica dell'Universit\`a di Pisa \& INFN Sezione di Pisa},
  \orgaddress{\street{Largo B. Pontecorvo 3}, \city{Pisa}, \postcode{56127}, \country{Italy}}}

\affil[3]{\orgdiv{Dipartimento di Fisica ``E. Pancini''}, \orgname{Universit\`a di Napoli Federico II},
  \orgaddress{\street{Complesso di Monte S. Angelo, via Cinthia}, \city{Napoli}, \postcode{80126}, \country{Italy}}}

\abstract{We consider a free-fermion chain undergoing dephasing, described by two different
  random-measurement protocols (unravelings): a quantum-state-diffusion and a quantum-jump one.
  Both protocols keep the state in a Slater-determinant form, allowing to address quite large system sizes.
  We find a {bifurcation} in the distribution of the measurement operators along the quantum trajectories,
  {that's to say, there is a point where the shape of this distribution changes} from unimodal to bimodal.
  The value of the measurement strength
  where {this phenomenon} occurs is similar for the two unravelings, but the distributions and the transition
  have different properties reflecting the symmetries of the two measurement protocols.
  We also consider the scaling with the system size of the inverse participation ratio of the Slater-determinant
  components and find a power-law scaling that marks a multifractal behaviour, in both unravelings and
  for any nonvanishing measurement strength.}

\maketitle

\section{Introduction}
\label{Sec:Introduction}

Entanglement~\cite{Nielsen, RevModPhys.81.865} plays an important role in the unitary dynamics
of many-body quantum systems in many different contexts: It marks quantum phase transitions~\cite{Amico_RMP};
It behaves differently in the dynamics of thermalizing, integrable and many-body localized
quantum systems~\cite{Alba_2017, Alba_2018, Singh_2016}; It even allows to detect the existence
of topological boundary modes~\cite{Mondal_2022,PhysRevB.101.085136,PhysRevB.105.085106,micallo}.
Recently, attention has been moved also to the behaviour of entanglement in situations beyond the unitary dynamics,
where the evolution of monitored systems is considered. The interplay between the intrinsic dynamics of the system
and that induced by the quantum measurement process can lead to a variety of scaling regimes for the asymptotic
entanglement entropy, giving rise to the so called entanglement transitions. 

In this framework, an extensive number of works has been focusing on local measurements (either discrete or continuous in time)
performed in monitored quantum 
circuits~\cite{Li2018, Chan2019, Skinner2019, Szyniszewski2019, Vasseur2021, Bao2021, Nahum2020, Chen2020,Li2019, Jian2020, Li2021, Szyniszewski2020, Turkeshi2020, Lunt2021, Sierant2022_B, Nahum2021, Zabalo2020, Sierant2022_A, Chiriaco2023, Klocke2023},
as well as in non-interacting~\cite{DeLuca2019,Nahum2020, Buchhold2021,Jian2022, Coppola2022, Fava2023, Poboiko2023, Jian2023, Merritt2023, Alberton2021, Turkeshi2021, Szyniszewski2022, Turkeshi2022, Piccitto2022, Piccitto2022e, Tirrito2022, Paviglianiti2023, chahine2023entanglement,Kells_2023}
and interacting~\cite{Lunt2020,Rossini2020, Tang2020, Fuji2020, Sierant2021, Doggen2022, Altland2022} Hamiltonian systems.
Moreover, there exists a deep connection between measurement-induced phases and the encoding/decoding properties of a quantum channel~\cite{Gullans2020_A, Gullans2020_B, Loio2023, Choi2020, Bao2020, Bao2021_A,Fidkowski2021, Bao2021_B, Barratt2022_A,Dehgani2023, Kelly2022}.
Situations where the dynamics is only induced by random measurements of non-local
string operators (measurement-only dynamics) have been also considered, finding different scaling regimes of the entanglement entropy, according to the statistics of the randomly measured operators, and the range and the nature of the strings~\cite{Ippoliti2021, Sriram2022}.

Among the various theoretical models of monitored quantum systems, considerable coverage has been dedicated to the dynamics
of fermionic Gaussian states, in the presence of quadratic Hamiltonians and Gaussian-preserving measurement
processes (see, e.g., Refs.~\cite{DeLuca2019, Lang2020, Alberton2021, Turkeshi2021, Turkeshi2022, Piccitto2022, Piccitto2022e, Coppola2022, Tirrito2022, Minato2022, Szyniszewski2022, Zerba2023, Paviglianiti2023, Poboiko2023,paviglianiti2023enhanced,Kells_2023,chahine2023entanglement}),
as they are amenable to an accurate numerical treatment up to relatively large sizes.
In this framework, for short-range Hamiltonians and local measurements, area-law (saturation to a finite value)
or logarithmic scaling of the asymptotic entanglement entropy with the system size have been reported.

In this paper we consider a free-fermion chain undergoing a dephasing Lindbladian, and describe it as an average
over random quantum trajectories in two different ways (unravelings).
Specifically, we employ either a {\it quantum-state-diffusion} (QSD) unraveling~\cite{DeLuca2019}
or a {\it quantum-jump} (QJ) unraveling. Both unravelings are chosen in such a way that the state is kept in a Gaussian
form (a Slater determinant), thus a numerical analysis of systems with up to $L=O(10^2)$ sites is easily affordable.

For this model, the behaviour of the asymptotic entanglement entropy averaged over the quantum trajectories
has been carefully scrutinized and is still under debate. After the pioneering work in Ref.~\cite{DeLuca2019},
where a saturation of the entanglement entropy with the system size (area-law) was predicted,
Refs.~\cite{Alberton2021,Ladewig2022} claimed the presence of a transition from a logarithmic increase
to an area-law behaviour.
More recently, Ref.~\cite{Poboiko2023} challenged this result, suggesting that there is only area law
and the transition is actually a crossover, due to the exponential growth with the inverse measurement strength
of the size where the entanglement entropy saturates.

Here we focus {instead on the distribution}
of the expectations of the measured operators along the quantum trajectories. 
Properties of similar distributions have already been studied in~\cite{passarelli2023postselectionfree,Tirrito2022,Russomanno2023_longrange}. {In general, the properties of this distribution are not related to the entanglement properties,
  although with an exception~\cite{passarelli2023postselectionfree}.} Comparing the behaviour of this distribution in the two unravelings, we observe differences
as well as similarities. In the QSD unraveling this distribution is symmetric around $n=1/2$, while in the QJ one
this symmetry is absent. This reflects the fact that, while QSD is invariant under particle-hole symmetry, QJ is not.
In both unravelings the distribution displays a {bifurcation}, moving from unimodal
to bimodal shape. {One can see this phenomenon already at finite size, being related to the frequent
  local measurements forcing the state to be locally similar to an eigenstate of the measurement operators,
  in analogy with the quantum Zeno effect~\cite{Zeno,PhysRevA.41.2295,PETROSKY1990109,ZenoSubspaces}.}
In the case of QSD, where the {bifurcation} occurs for a measurement strength $\gamma_{\rm QSD}^\star \approx 0.2$,
the two maxima in the bimodal phase are symmetric and stem continuously at the transition point, in a way {formally} reminiscent of
the Landau picture of second-order phase transitions. At the {bifurcation} point, the single maximum bifurcates
into two maxima, with a discontinuity in the derivative analogous to the pitchfork {bifurcation}
in classical dynamical systems~\cite{cross:book,strogatz:book,gold:book}. In the QJ case, the {bifurcation}
occurs at $\gamma_{\rm QSD}^\star \approx 0.23$. In contrast with the QSD case, here the maxima are not symmetric,
and the smaller one appears discontinuously at the {bifurcation} point, in a way {formally} reminiscent of the mean-field analysis
of first-order phase transitions.

The {bifurcation} originates from the interplay between the unitary dynamics and the measurement operations.
In particular, when the measurement strength is large, the second prevail and the state is expected to be similar
to a product state. Consistently with that, in both cases, for large measurement strength, the distributions show
strong maxima near $n=0$ and $n=1$, marking that the state is near to a separable one. This also agrees with
what is known for the asymptotic entanglement entropy~\cite{DeLuca2019,Alberton2021,Ladewig2022,Poboiko2023},
that for QSD tends to 0 for large measurement strengths.

Finally we study the localization of the state, to probe whether the {bifurcation}
corresponds to a delocalization-localization transition or not~\footnote{Localization in monitored systems has also been considered in Refs.~\cite{chahine2023entanglement,PhysRevResearch.5.033174,patrick2023enhanced}.}. In order to do that, we first note that, for both unravelings,
the state is a Slater determinant. We consider the components of this determinant and evaluate
the inverse participation ratio (IPR) of these components in the space basis.
This is a standard probe for localization, used for instance in studies of Anderson localization~\cite{Edwards_JPC72}.
We find no sharp transition, but we see that the IPR scales with the system size $L$ as a power law $\sim L^{-\alpha}$
with $0<\alpha<1$. For both unravelings, $\alpha$ depends smoothly on $\gamma$ {similarly to what happens in the two-dimensional case~\cite{chahine2023entanglement}.}
This is a mark of multifractal behaviour of the components of the Slater determinant~\cite{PhysRevB.66.033109,Mirlin}: {The system is anomalously delocalized and never achieves perfect localization, as confirmed by results on the conductivity~\cite{PhysRevResearch.5.033174}}.
Such multifractal behaviour also occurs at the transition between extended and Anderson-localized phases,
so we can say that our model is always in a critical Anderson regime.

The paper is organized as follows.  In Sec.~\ref{Sec:Model} we present the model Hamiltonian, its diagonalization
and its symmetries. In Sec.~\ref{Sec:Unravelings} we describe the two unraveling protocols that, on average, give rise
to the same Lindblad equation. In Sec.~\ref{Sec:Methods} we show that for both unravelings the state along a trajectory
is always in a Slater determinant form. In Sec.~\ref{Sec:Data} we compare the two unravelings and show that,
although the asymptotic entanglement entropy behaves similarly, the distributions of the expectations
of the measurement operators have different symmetries, but show both a {bifurcation} (Sec.~\ref{distrib:sec});
we also discuss the scaling of the IPR averaged over disorder and its multifractal behaviour (Sec.~\ref{ipr:sec}).
Finally, in Sec.~\ref{Sec:Conclusions} we draw our conclusions.

\section{Model Hamiltonian}
\label{Sec:Model}

We consider a model of spinless free fermions on a one-dimensional lattice with $L$ sites,
described by the Hamiltonian
\begin{equation}
  \hat{H} = \frac{\lambda}{2}\sum_j\left(\opcdag{j}\opc{j+1}+\opcdag{j+1}\opc{j}\right),
  \label{eq:Ham0}
\end{equation}
where $\hat c^{(\dagger)}_j$ are anticommuting fermionic operators acting
on the $j$-th site~\footnote{In what follows, unless specified, summations run from $1$ to $L$
and we assume periodic boundary conditions.
We also fix $\lambda = 1$ as the energy scale of the system and work in units of $\hbar=1$.}.
After defining the  vector $\hat \Psi \equiv (\hat c_1, \dots, \hat c_L)^T$,
the Hamiltonian~\eqref{eq:Ham0} can be written as
\begin{equation}
  \hat H =  \hat \Psi^\dagger \mathbb{H} \hat \Psi,
  \label{eq:ham}
\end{equation}
where we introduced the $L \times L$ matrix $\mathbb{H}$ such that
\begin{equation}
   \mathbb{H}_{i, j} = \frac{\lambda}{2} (\delta_{i+1, j}+ \delta_{i-1, j}).
  \label{eq:bdg}
\end{equation}

A generic fermionic quadratic model as the one in Eq.~\eqref{eq:ham} can be cast
in a more transparent form by defining the fermionic operators
\begin{equation}
  \hat \gamma_k = \sum_j  \mathbb{U}_{kj}\hat c_j,
  \label{eq:gamma_f}
\end{equation}
such that
\begin{equation}
  \hat \Phi \equiv \big(\hat \gamma_1, \dots, \hat \gamma_L\big)^T
  = \mathbb{U}\hat \Psi ,
\end{equation}
where $\mathbb{U}$
denotes the unitary transformation which diagonalizes this matrix $\mathbb{H}$:
\begin{equation}
  \mathbb{U}^\dagger \, \mathbb{H} \, \mathbb{U} = \text{diag}\big(\omega_k\big), \quad (k=1,\ldots, L).
\end{equation}
Thus, we finally get:
\begin{equation}
  \hat H = \sum_k \omega_k \hat \gamma_k^\dagger \hat \gamma_k\,,
  \label{eq:Hfree}
\end{equation}
which represents a model of free $\hat \gamma_k$-fermionic particles
with dispersion relation $\omega_k \geq 0$.

The eigenstates of $\hat H$ can be put in a simple Gaussian form (Slater determinants)
and are fully characterized by the two-point correlation matrix of $\hat c$-fermions,
$C_{ij} = \braket{\hat c_i^\dagger \hat c_j}$ (see Sec.~\ref{Sec:Methods}).
This makes such kind of quadratic fermionic systems particularly suited to numerical computations
up to large sizes $L \sim O(10^3)$. 
It is also worth noticing that Gaussianity of states is preserved by the application of any operator
that can be written as the exponential of an other operator that is quadratic
in the fermionic operators (see, e.g., Refs.~\cite{Bravyi2005, DiVincenzo2005, DeLuca2019, Piccitto2023}).

We finally remark that the Hamiltonian in Eq.~\eqref{eq:Ham0} conserves the total fermion number
$\hat{N} = \sum_j\opcdag{j}\opc{j}$ and is invariant under the particle-hole transformation, the one that exchanges
every $\opcdag{j}$ with the corresponding $\opc{j}$, and thus exchanges every $\hat{n}_j$ with $\hat{1}-\hat{n}_j$.

\section{Monitored dynamics}
\label{Sec:Unravelings}

To characterize the effects of measurements on the unitary quantum dynamics generated
by the Hamiltonian in Eq.~\eqref{eq:Ham0}, we construct a suitable stochastic Schr\"odinger equation,
whose exact form depends on the kind of measurement protocol (unraveling) one wants to study~\cite{Plenio1998, Daley2014}. The measurements provide noise and on each stochastic realization of the noise the state evolves along a so-called quantum trajectory.
In what follows, we focus on two distinct unravelings that give rise to the same
master equation for the mean density matrix of the system, once averaged over the stochastic realizations.

\subsection{Quantum state diffusion}
\label{sec:QSD}

In the first measurement protocol, we assume the number $\langle \hat n_j \rangle_t$ of fermions on the $j$-th site
(with $\hat n_j = \hat c^\dagger_j \hat c_j$) to be continuously measured with a rate $\gamma$.
This generates a QSD dynamics, that is, a collection of Wiener processes.
For small time evolution steps $\delta t$, it can be cast in the following form~\cite{DeLuca2019}:
\begin{equation}\label{step:eqn}
  \ket{\psi(t+\delta t)} \! \propto \! e^{\sum_j \!\! \big[\delta W^j_t + (2 \braket{\hat n_j}_t -1)\gamma \delta t\big]\hat n_j} 
  e^{-i\hat H \delta t} \! \ket{\psi(t)}\!,
\end{equation}
neglecting normalization constants and assuming $\braket{\cdot}_t \equiv \braket{\psi(t)|\cdot |\psi(t)}$.
The $\delta W_t^i$ variables are normally distributed with zero mean and variance $\gamma \, \delta t$.

We emphasize that this unraveling is invariant under the particle-hole transformation. Indeed, by exchanging
$\hat{n}_j \leftrightarrow \hat{1}-\hat{n}_j$ in Eq.~\eqref{step:eqn}, one gets back the same equation
with an immaterial multiplying factor in front of the state and the sign of $\delta W_t^j$ changed.
This different sign is immaterial, since the $\delta W_t^j$ are Gaussian variables with zero mean, and all the moments
of any distribution over the randomness are left unchanged by the transformation. 

\subsection{Quantum jumps}
\label{sec:QJ}

Secondly, we consider an unraveling based on occasional yet abrupt measurements of the local operators $\hat m_\ell = \hat 1 + \hat n_\ell$,
where $\hat 1$ denotes the identity operator. We assume that these measurements can occur on any site $\ell$
with a probability $p_\ell \in [0,1]$ that we define below.
This kind of measurements are modeled by a QJ dynamics that describes the evolution of the system for discrete
time steps $\delta t$. At each step one of the following operations on the state can randomly occur: 
\begin{enumerate}
\item With probability $p_\ell = \gamma \braket{\hat m^2_\ell}_t \delta t= \gamma (1+3\braket{\hat n_\ell}_t)\delta t$,
  where $\gamma$ denotes the measurement rate, we project the system state with one of the $L$ measurement
  operators~\footnote{To maintain convergence of the QJ protocol, one should ensure the condition
  $\sum_j p_j = \gamma \, \delta t \sum_j \langle \hat m_j^2 \rangle_t \ll 1$.}:
  \begin{equation}\label{acuna:eqn}
    \ket{\psi(t + \delta t)} = \frac{\hat m_\ell \ket{\psi(t)}}{||\hat m_\ell \ket{\psi(t)}||};
  \end{equation}
\item With probability $\bar p = 1 - \sum_j p_j$ we evolve with a non-Hermitian effective Hamiltonian $\hat H^\text{eff}$
  \begin{equation}\label{matata:eqn}
    \ket{\psi(t+\delta t)} = e^{-i \hat H^\text{eff} \delta t} \ket{\psi(t)},	
  \end{equation}
  where $\hat H^\text{eff} = \hat H - \frac{3}{2}i\gamma \sum_j \hat n_j$.
\end{enumerate}
Since for spinless fermions $\hat 1 + \hat{n}_j = \nep^{\hat{n}_j\ln 2}$ (at most one fermion per site
is allowed), it is immediate to see that this kind of dynamics can be implemented
by preserving Gaussianity~\cite{Piccitto2022, Piccitto2022e}.
Moreover, as one can easily check, it is not invariant under particle-hole transformation.

\subsection{Average over trajectories}
\label{sec:avg}

When averaged over the stochastic realizations, in the limit $\delta t\to 0$, both QSD and QJ unravelings
lead to the same Lindblad master equation
for the mean density matrix $\rho_t = \overline{\ket{\psi(t)}\bra{\psi(t)}}$,
where the overline indicates the ensemble average,~\cite{Petruccione}
\begin{equation}
  \partial_t\rho_t \!= -i [\hat{H}, \rho_t]
  + \gamma \!\sum_j \!\Big( \hat{m}_j \, \rho_t \, \hat{m}_j - \tfrac{1}{2} \{ \hat{m}_j,\rho_t \} \Big).
  \label{eq:lind}
\end{equation}
In this equation, $\hat m_j$ denote the (Hermitian) measurement operators and $\gamma$ quantifies
the strength of the coupling between the system and the measurement apparatus.
Note that in the QSD protocol (Sec,~\ref{sec:QSD}) we set $\hat m_j \equiv \hat n_j$,
while in the QJ protocol (Sec.~\ref{sec:QJ}) we set $\hat m_j = \hat 1 + \hat n_j$.
In fact, substituting $\hat m_j \to \hat 1 + \hat n_j$, one recovers the same Lindblad equation
for the measurement operator $\hat n_j$.

In other words, the two stochastic dynamical protocols above are unravelings
of the same Lindblad master equation~\eqref{eq:lind}~\cite{Piccitto2022, Piccitto2022e}.
As a consequence, any linear function of the density matrix $\rho_t$ must be independent of the chosen unraveling. 
We mention, for example, the expectation value of a generic observable $\hat O$,
\begin{equation}
  \overline{\braket{\hat O}_t} = \text{Tr} \big[\hat O \rho_t\big].
  \label{eq:Oavg}
\end{equation}

Finally we point out that, independently of $\gamma$, the dynamics described above preserves the total number
of fermions $\hat N = \sum_j \hat n_j$. As a consequence, one can work in a subspace of the full Hilbert space
with a fixed filling. In what follows we work in the half-filling subspace and fix $N\equiv \braket{\hat N}  = L/2$.

\section{Methods}
\label{Sec:Methods}

To reduce the numerical complexity of the problem, we exploit the conservation of the number $N$ of fermions,
as stated above. We choose as initial condition the N{\'e}el state with $N=L/2$ particles:
\begin{equation}
  \ket{\psi(0)} = \prod_{j=1}^{L/2} \hat c_{2j}^\dagger \ket{\Omega},
\end{equation}
where $\ket{\Omega}$ is the vacuum of $\hat c$-fermions, such that $\hat c_j \ket{\Omega} = 0$, $\forall j=1,\ldots,L$.
Then, we let it evolve according to one of the two measurement protocols described before, up to time $t$.
The time evolved state $\ket{\psi(t)}$ can be cast in the form of a generic Gaussian state.
The full information of such state is contained in a $L \times N$ matrix $U_t$, defined by
\begin{equation}
  \ket{\psi(t)} := \prod_{k=1}^N \left[ \, \sum_{j=1}^L \big[ U_t \big]_{jk} \, \hat c_j^\dagger\right] \ket{\Omega},
  \label{eq:Psi_t}
\end{equation}
such that $U_t^\dagger U_t = \mathbb{I}_{N \times N}$ and $U_t \, U_t^\dagger = C(t)$,
being $C(t)$ the two-point correlation matrix with elements $C_{ij}(t) = \braket{\hat c_i^\dagger \hat c_j}_t$.
The state Eq.~\eqref{eq:Psi_t} is in a Slater-determinant form~\cite{mermin:book}.

As explained in Ref.~\cite{DeLuca2019}, for the QSD protocol, the dynamics can be split in two steps
accounting for the unitary and the dissipative one. 
Note that, in general, the operators generating these steps do not commute, therefore this approach
is based on a Trotterization of the full dynamics in the limit of small time steps $\delta t$.
The dynamical step thus reads 
\begin{equation}
  V_{t+\delta t} = Me^{-i \, \mathbb{H} \, \delta t} U_t,
  \label{eq:Trotter}
\end{equation}
with 
\begin{equation}
  M_{jk}= \delta_{jk}e^{\delta W_t^j + (2\braket{\hat n_j}_t -1)\gamma \delta t}.
\end{equation}
Since the whole dynamics does not preserve the norm of the state, to restore the unitarity,
at the end of the step in Eq.~\eqref{eq:Trotter} we have to perform a QR-decomposition
\begin{equation}
  V_{t+\delta t} = \begin{bmatrix}Q_1 & Q_2 \end{bmatrix}\begin{bmatrix} R_1 \\ 0 \end{bmatrix} = Q_1 R_1,
  \label{eq:QR}
\end{equation}
where $R_1$ is a $N \times N$ upper triangular matrix, $Q_1$ is $L\times N$ and $Q_2$ is $L \times (L - N)$. 
The time evolved state is then characterized by $U_{t+\delta t} \equiv Q_1$.

For the QJ protocol, we exploit a similar procedure.
The matrix after the projective step Eq.~\eqref{acuna:eqn} on the $\ell$-th site is obtained from $V_{t+\delta t} = TU_t$ with 
\begin{equation}
  T_{ij}= 2 \delta_{ij} \, \delta_{i\ell} \, \delta_{j\ell},
\end{equation}
while the state after the non-Hermitian step Eq.~\eqref{matata:eqn} is obtained by applying
$V_{t+\delta t} = e^{-i \, \mathbb{H}^\text{eff} \, \delta t} U_t$,
\begin{equation}
  \mathbb{H}_{ij}^\text{eff} = \mathbb{H}_{ij} - \tfrac{3}{2} i \gamma \delta_{ij}.
\end{equation}
Even in this case, at the end of each step, we have to restore the unitarity of $V_{t+\delta t}$
by performing a QR-decomposition [see the above Eq.~\eqref{eq:QR} and the related discussion].

\section{Comparison of the two unravelings}
\label{Sec:Data}

In this section we discuss the differences in the dynamical behaviour of the system, under the action of the two unravelings described in Sec.~\ref{Sec:Unravelings}.
In particular we focus on:
(i) the distribution of the expectation values of the local number operator,
once this is measured in time over the various trajectories (Sec.~\ref{distrib:sec});
(ii) the localization properties of the states averaged along the trajectories (Sec.~\ref{ipr:sec}).
In all the following we are going to perform evolutions up to a time $t_f = 10^3$, and choose $\delta t = 0.05$ for QSD and $\delta t = 0.16/L$ for QJ, in order to ensure convergence.

\subsection{Distribution of $\langle \hat n_j \rangle_t$}
\label{distrib:sec}

\begin{figure}
  \centering
  \includegraphics[width=88mm]{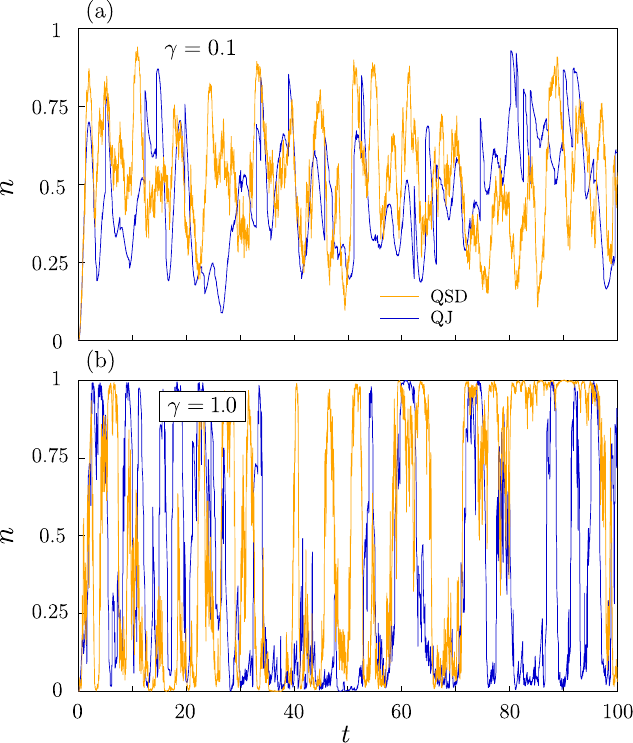}
  \caption{Temporal behaviour of $n \equiv \braket{\hat n_j}_t$ for the two unravelings (orange and blue curves
    for QSD and QJ, respectively) for $\gamma = 0.1$ (a) and $\gamma = 1.0$ (b).
    In both cases, curves oscillate around the mean value $n = 0.5$. However, for small measurement
    rates they remain concentrated around the mean value, while for large $\gamma$ they shift close to the
    extremes.}
  \label{fig:nj_single_realization}
\end{figure}

We first consider the statistics of the expectation values {$\braket{\hat n_j}_t\equiv\braket{\psi_t|\hat{n}_j|\psi_t}$} detected
along the quantum trajectories in the different unravelings.
Because of translation invariance {of the average dynamics}, in the long-time limit {the statistics of this expectation value} does not depend
on the site $j$, therefore we can drop the explicit dependence. To further simplify the notation,
hereafter we also drop the time dependence and pose $n \equiv \braket{\hat n_j}_t$.
As discussed in Sec.~\ref{sec:avg}, because of the linearity of the trace operation,
the ensemble average over the two stochastic processes (QSD and QJ) should coincide [see Eq.~\eqref{eq:Oavg}].
We checked this numerically and indeed we found that $\overline{n} = 1/2$,
independently of the unraveling: This agrees with the fact that we work in the half-filling subspace.

\begin{figure}
  \centering
  \includegraphics[width=88mm]{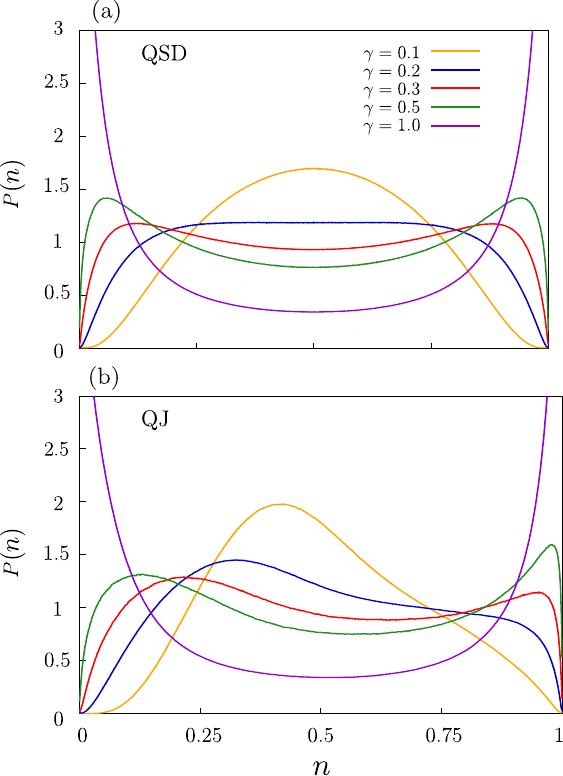}
  \caption{Distributions of $n$ along a time trajectory, averaged over $N_r = 80$ noise
    realizations for QSD (panel a) and QJ (panel b) protocols. The various colors refer to different values of $\gamma$.
    These plots have been obtained for $L=128$, but the curves are almost independent of the system size
    (see Fig.~\ref{fig:dos_L}).}
  \label{distri:fig}
\end{figure}

{Let us start showing some examples of $\braket{\hat n_j}_t$ versus $t$ along single trajectories, for a single noise realization.}
Examples of such trajectories are plotted in Fig.~\ref{fig:nj_single_realization} for (a) $\gamma = 0.1$ 
and (b) $\gamma = 1.0$. 
The colors refer to the two stochastic processes, QSD (orange) and QJ (blue).
For both unravelings, the trajectories fluctuate around the ensemble average value
$\overline{n} = 1/2$,
however the nature of these oscillations depends on $\gamma$. In fact, while for small measurement rates they stay close
around $0.5$ (panel a), for larger $\gamma$ values they shift and spend more time near the extremal values $0$ and $1$ (panel b). 

This can be better appreciated in Fig.~\ref{distri:fig}, where we show the probability distribution $P(n)$ 
along a time trajectory for different values of $\gamma$ and for both QSD (panel a) and QJ (panel b)
unraveling protocols. {We evaluate this distribution as the histogram of $\braket{\psi_t|\hat{n}_j|\psi_t}$ along the trajectory, averaging over sites $j$ and noise realizations, in order to get smoother curves. Furthermore, we evaluate each distribution over a set of trajectories lasting a time $t_f$, large enough that all transients have vanished.}
%
The main emerging feature is that, for increasing $\gamma$, the distribution crosses from a unimodal
to a bimodal character, reflecting the differences in the trajectories discussed in Fig.~\ref{fig:nj_single_realization}.
We refer to this behaviour as a ``quantum {bifurcation}''~\cite{Russomanno2023_longrange}.

Looking at the shape of $P(n)$, we can see some differences between the two unravelings. 
In fact, because of the particle-hole symmetry (see Sec.~\ref{sec:QSD}), in the QSD dynamics the distribution is symmetric
around the ensemble average value $\overline{n}=1/2$.
This behaviour is reminiscent of the minima of the free energy in the Landau model.
In contrast, for the QJ protocol we do not observe the same symmetry, consistently with the fact that
such unraveling breaks the particle-hole symmetry (see Sec.~\ref{sec:QJ}), but we can still identify a crossover
from a unimodal asymmetric distribution to a bimodal asymmetric one, where a second local maximum appears.
In any case, we carefully checked that, even in this latter case, the average value
$\overline{n} = 1/2$ is consistent with the half-filling assumption
and the conservation of particle number. 

\begin{figure}
  \centering
  \includegraphics[width=88mm]{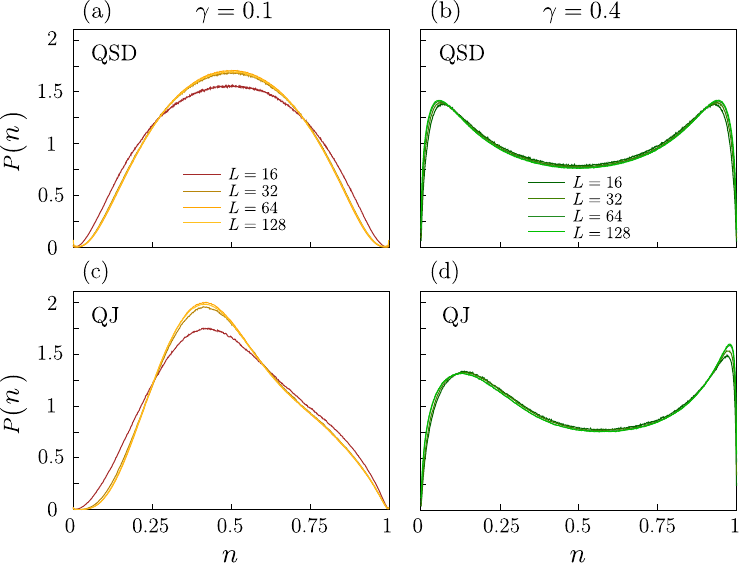}
  \caption{Distributions of $n$ along a time trajectory, averaged over $N_r = 80$ noise
    realizations for QSD (top panels) and QJ (bottom panels) protocols, for $\gamma = 0.1$ (left column)
    and $\gamma = 0.4$ (right column). The various colors refer to different system sizes.}
  \label{fig:dos_L}
\end{figure}

We should also emphasize that the numerical data reported in Fig.~\ref{distri:fig} are for a system with $L=128$ sites.
{However they are almost independent of the system size, as explicitly shown in Fig.~\ref{fig:dos_L}, in analogy with the quantum Zeno effect~\cite{Zeno, PhysRevA.41.2295, PETROSKY1990109, ZenoSubspaces}.}
As expected, a more evident dependence on $L$ emerges for smaller values of $\gamma$ (left panels, $\gamma=0.1$),
while for larger measurement strengths (right panels, $\gamma=0.4$) differences in the curves from $L=16$ to $L=128$
become hardly visible on the scale of the figure.

\begin{figure}
  \centering
  \includegraphics[width=0.45\textwidth]{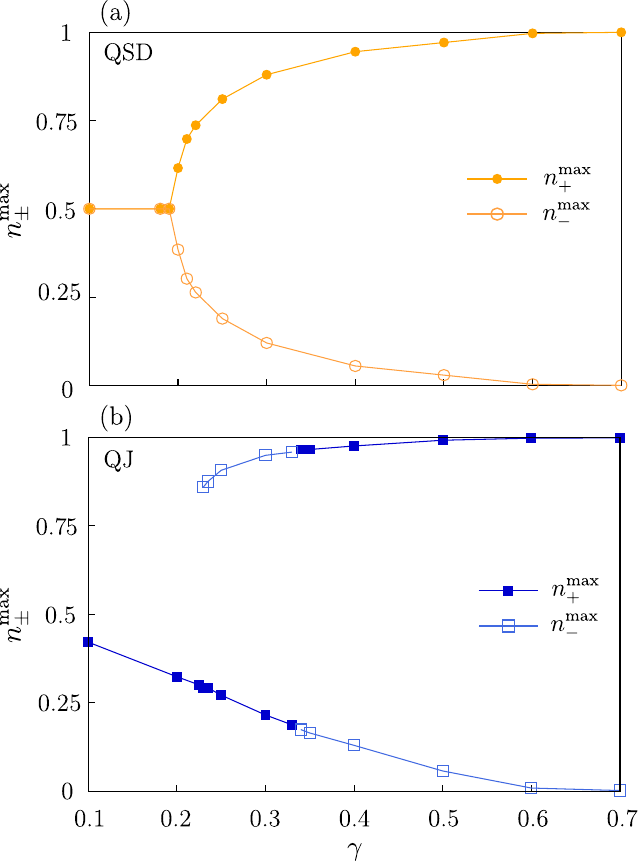}
  \caption{Position of the local maxima of $P\big(n\big)$ vs $\gamma$
    for QSD (panel a) and QJ (panel b) protocols. For QSD the two maximua stem continuously
    and symmetrically at the {bifurcation} point $\gamma_{\rm QSD}^\star \approx 0.2$, with a discontinuity in the derivative
    with respect to $\gamma$.
    For QJ the maxima are asymmetric: the secondary one appears discontinuously at $\gamma_{\rm QJ}^\star \approx 0.23$,
    while the global and the secondary one swap each other at $\gamma \approx 0.35$.
    Numerical parameters: $L = 128,\,N_r\geq 48$.}
  \label{distrim:fig}
\end{figure}

To gain further insight on the properties of the {bifurcation}, in Fig.~\ref{distrim:fig} we show
the position of the two local maxima of the distribution,
$n^{\rm max}_-$ and $n^{\rm max}_+$, defined such that $P(n^{\rm max}_+) \geq P(n^{\rm max}_-)$.
In Fig.~\ref{distrim:fig}(a) we consider the case of the QSD protocol: We see that, at the {bifurcation} point
$\gamma_{\rm QSD}^\star\approx 0.2$, the two maxima stem symmetrically (i.e., $n^{\rm max}_+ - \overline{n} = |n^{\rm max}_- - \overline{n}|$) and continuously  from the single maximum,
and the derivative with respect to $\gamma$ is discontinuous at the {bifurcation}.
This behaviour is {formally} similar to the mean-field analysis of a second-order phase transition (the maxima here
correspond to minima of the free energy there), or a pitchfork {bifurcation}
in classical dynamical systems~\cite{cross:book, strogatz:book, gold:book}.
In Fig.~\ref{distrim:fig}(b) we consider the case of the QJ protocol: The maxima are not symmetric, and we call the one
corresponding to the smaller value as ``secondary maximum''. Here the {bifurcation} occurs in a different way
compared to the QSD case, because at the {bifurcation} point $\gamma_{\rm QJ}^\star \approx 0.23$ the secondary maximum
appears discontinuously. For $\tilde \gamma \approx 0.35$ the global and the secondary maxima swap with each other.
This behaviour is {formally} similar to what happens in a first-order phase transition, with the maxima of our distribution
corresponding to the minima of the free energy in that case.

We notice that, for both unravelings, at $\gamma \approx 0.4$ the position of the global maximum
approaches the value $n_+^\text{max}=1$. This is consistent with the fact that, for large values of $\gamma$,
the measurement prevails and the state of the system is close to a product one, as the results on the asymptotic
entanglement entropy for the QSD unraveling confirm~\cite{DeLuca2019, Alberton2021, Ladewig2022, Poboiko2023}.
We have checked that the entanglement entropy behaves similarly also for the QJ case (such kind of similarity
in the entanglement behaviour for different unravelings of the same Lindblad dynamics contrasts with what has been observed
with the additional presence of nearest-neighbour coherent pairing terms, which modify the symmetry properties
of the Hamiltonian~\cite{Piccitto2022,Piccitto2022e}).
Nevertheless we stress that the entanglement transition and the {bifurcation} are different
and unrelated phenomena~\cite{Russomanno2023_longrange}, thus the above analysis of the probability distributions
is not expected to provide information on the entanglement behaviour.

\subsection{Inverse participation ratio}\label{ipr:sec}
  
A related question is whether the qualitative change in the behaviour of the distributions
of $\langle \hat n_j \rangle_t$ could be related with some change in the space-delocalization properties of the state. 
To this purpose, we study the time averaged IPR of the components of the Slater determinant, defined as
\begin{equation}
  \overline{\text{IPR}} = \frac{1}{t_f} \int_0^{t_f} \frac{2}{L} \sum_{\mu=1}^N \text{IPR}_\mu(t) dt,
\end{equation}
with 
\begin{equation}
  \text{IPR}_\mu(t) = \sum_{j=1}^L |U_{j\mu}(t)|^4.
\end{equation}
In general, the IPR is such that $1/L \le \sum_\mu \text{IPR}_\mu \le 1$,
where the two bounds refer to a perfectly delocalized and a localized state, respectively. 

\begin{figure}[!t]
  \centering
  \includegraphics[width=88mm]{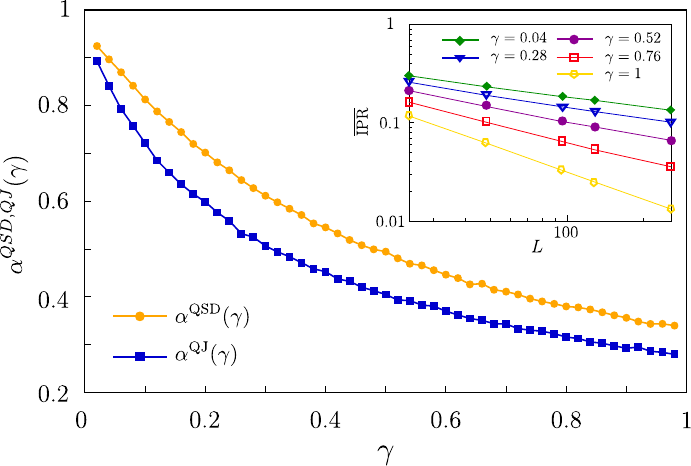}
  \caption{The factors $\alpha^\text{\rm QSD}$ (orange dots) and $\alpha^\text{\rm QJ}$ (blue squares)
    vs $\gamma$, as obtained by a numerical fit of the IPR data with a power-law as a function of $L$,
    cf. Eq.~\eqref{eq:IPRfit} (see the inset in log-log scale, for the QSD protocol).}
    \label{slope:fig}
\end{figure}

Let us first focus on the QSD unraveling. To characterize the localization properties when varying
the measurement strength, we numerically evaluate the IPR as a function of the system size
and for different values of $\gamma$, as shown in the inset of Fig.~\ref{slope:fig}.
Different lines refer to various values of $\gamma$, from $\gamma = 0.04$ to $\gamma = 1$,
with steps of $\delta \gamma = 0.04$.
Qualitatively similar results have been obtained for the QJ protocol (for this reason,
we decided to show only data for the QSD protocol).
We then fitted the data with a power law
\begin{equation}
  \overline{\rm IPR}\sim L^{-\alpha(\gamma)}.
  \label{eq:IPRfit}
\end{equation}
and found that both unravelings exhibit a similar scaling with $L$, but with a power-law exponent
$\alpha(\gamma)$ that depends on the considered measurement protocol (main frame of Fig.~\ref{slope:fig},
where the orange dots refers to QSD, while the blue squares to QJ).
In both cases, the single-particle states are perfectly delocalized ($\alpha = 1$) only in the $\gamma\to 0$ limit.
For $\gamma > 0$, the scaling exponents deviate from $1$ and differ from each other.
This behaviour is typical of anomalous delocalization of states with multifractal properties~\cite{PhysRevB.66.033109,Mirlin}.
The fact that we do not observe discontinuities for $\alpha$ as a function of $\gamma$,
seems to indicate that delocalization properties are independent of the possible presence of
an entanglement transition. {In particular, $\alpha$ continuously decreases with increasing $\gamma$,
 marking that the system becomes less and less delocalized, but never attains localization. This result fits with the findings
 of~\cite{PhysRevResearch.5.033174}, where the authors see that the conductivity decreases as a power law with $\gamma$ but never vanishes.}

\section{Conclusions}
\label{Sec:Conclusions}

In conclusion we studied the dynamics of a free-fermion chain under dephasing, considering two protocols
of random measurements that, once averaged over the quantum trajectories, provide the same Lindbladian:
From one side we considered the QSD unraveling that preserves the particle-hole symmetry,
and from the other side a QJ unraveling that breaks such symmetry.
In both protocols the state of the system can be always cast as a Slater determinant,
allowing us to analyze quite large sizes.

We focused on the expectations of the measured operators along the trajectories and studied the properties
of their distributions. When the measurement strength $\gamma$ lies below a certain threshold,
the distributions are unimodal, while above this threshold they become bimodal, giving rise to a {bifurcation} transition.
This {bifurcation} threshold is located at $\gamma_{\rm QSD}^\star \approx 0.2$ for QSD and at $\gamma_{\rm QJ}^\star \approx 0.23$ for QJ,
and QSD preserves particle-hole symmetry
providing distributions symmetric around $n=1/2$, while QJ breaks this symmetry and gives rise to asymmetric distributions.
For QSD, two symmetric maxima appear at the {bifurcation}, in {formal} analogy with second-order phase transitions.
The two maxima stem from the single one continuously, with a discontinuity in the derivative in $\gamma$, as occurring
for the pitchfork {bifurcation} in classical dynamical systems.
On the other hand, in the QJ case the two maxima are asymmetric and the secondary one appears discontinuously,
as occurring in first-order phase transitions.
Quite curiously, the distributions display a very weak dependence on the system size.

Finally we analyzed the behaviour of the IPR of the components of the Slater determinant,
averaged over time and randomness. We considered the scaling of this quantity with the system size $L$,
to understand the space localization properties of the state, in analogy with what is done
in Anderson-localization problems. We always find a power-law scaling of the form $\sim L^{-\alpha}$,
where the exponent $\alpha$ equals 1 for $\gamma\to 0$, marking perfect delocalization.
For $\gamma > 0$, we find that $\alpha$ depends continuously on $\gamma$ and for both unravelings $0<\alpha<1$,
marking a multifractal behaviour of the Slater-determinant components. This is what happens at the
localization-delocalization transition in Anderson-localization problems,
so we can say that our model is always in an Anderson critical phase.

Perspectives of future work focus on the study of properties of the measurement-operator distributions
and of the IPR in nonintegrable monitored models~\cite{xing2023interactions,Tang2020,PhysRevLett.127.140601},
in order to understand the possible existence of quantum {{bifurcation}s} or  localization transitions in these cases.

\backmatter

\bmhead{Acknowledgments}

We acknowledge fruitful discussions with M.~Fava. G.~P. and A.~R. acknowledge financial support from PNRR MUR Project PE0000023-NQSTI. A.~R. acknowledges computational resources from MUR, PON ``Ricerca e Innovazione 2014-2020'', under Grant No.~``PIR01\_00011--(I.Bi.S.Co.)''.

\section*{Declarations}

The authors declare they have no financial interests nor conflicts of interest to this work.


\end{document}